\documentclass[a4paper,10pt]{article}
\usepackage{amsmath,amssymb,mathtools}  
\usepackage{soul}                       
\usepackage[usenames,dvipsnames]{xcolor}
\usepackage[raggedright]{titlesec}
\usepackage{scrextend}
\deffootnote{2em}{0em}{\thefootnotemark\quad}
\usepackage{multicol}
\usepackage[left=0.75in,right=0.75in,top=1.0in,bottom=1.25in]{geometry}
\usepackage{hyperref}
\usepackage{xcolor}

\hypersetup{
  colorlinks   = true, 
  urlcolor     = blue, 
  linkcolor    = blue, 
  citecolor   = blue 
}

\setstcolor{red}    
\setulcolor{red}    
\allowdisplaybreaks 

\let\oldabstract\abstract
\let\oldendabstract\endabstract
\makeatletter
\renewenvironment{abstract}
{%
               {\list{}{\addtolength{\leftmargin}{4em} 
                        \listparindent 1.5em%
                        \itemindent    \listparindent%
                        \rightmargin   \leftmargin%
                        \parsep        \z@ \@plus\p@}%
                \item\relax}%
               {\endlist}%
\oldabstract}
{\oldendabstract}
\makeatother

\title{\LARGE\textbf{\textsf{Cognitive Reflection Test and the Polarizing Force-Identification Questions in the FCI}}}

\author{\normalsize Allan L. Alinea}

\date{}

\begin{document}

\maketitle
\vspace{-2.25em}
\noindent
\begin{center}
{\small Institute of Mathematical Sciences and Physics,\\ University of the Philippines Los Ba\~nos,\\College, Los Ba\~nos, Laguna 4031 Philippines, alalinea@up.edu.ph}
\end{center}

\bigskip
\begin{abstract}
	\noindent{The set of polarizing force-identification (PFI) questions in the FCI consists of six items all basically asking only one question: the set of forces acting on a given body. Although it may sound trivial, these questions are among the most challenging in the FCI. In this work involving 163 students, we investigate the correlation between student performance on the set of PFI questions and the Cognitive Reflection Test. We find that for both scores in the FCI as a whole and in the PFI questions, the range of values of the Pearson coefficient at 95\% confidence interval, is suggestive that cognitive reflection may be one of the contributing factors in the student performance in the FCI. This is consistent with the idea that high level of cognitive reflection may help in eliminating seemingly valid choices (misconceptions) in the FCI that are intuitive from everyday experience or ``common sense'' but otherwise misleading. The ability to activate System 2 in Dual Process Theory, whether from System 1 or right after reading a physics problem, may contribute in narrowing down the set of prospective valid answers in a given physics problem. Complementary to cognitive reflection are other factors associated with deep understanding of physics whose effects are expected to become more evident with the level of difficulty of a set of physics problems. Given two students with the same level of cognitive reflection, the one with deeper understanding of physics is more likely to get the correct answer. In our analysis, the range of correlation coefficient for the set of PFI questions is downshifted with respect to that for the FCI as a whole. This may be attributed to the more challenging nature of the latter compared to a significant fraction of the remaining questions in the former.}
\bigskip
\\
\noindent
{\small\textbf{keywords:} \textit{Polarizing Force-Identification Questions; Cognitive Reflection Test; Force\\\phantom{Concept Inv}Concept Inventory; Dual Process Theory; Physics Education}}	
\bigskip
\\
\noindent
{\textit{accepted for publication: European Journal of Physics}}	
\end{abstract}

\bigskip
\begin{multicols}{2}
\section{Introduction}
In spite of the enormity and complexity of possible information configurations that it can process, it appears that the way human mind thinks or decides on a given problem or situation, can often be simplified into merely two systems. Dual process theory (DPT) in psychology distinguishes these two systems as \textit{intuitive} (`System 1'), one that is considered fast and autonomous, and \textit{analytic} (`System 2'), the other one that is considered as deliberative and slow; see e.g., Refs. \cite{Neys, Stanovich, Kahneman}. System 1 allows us to make quick decisions with minimal strain on our mental resources (e.g., recognizing a friend in a classroom). On the other hand, System 2 enables us to solve problems where higher-order thinking is required (e.g., solving a $10 \times 10$ \textit{sudoku}). 

In dealing with different situations, it is helpful to efficiently choose the appropriate process to use. More properly, with System 1 often proceeding ``unconsciously,'' it is important to be cognizant of the need to activate System 2 on top of System 1. The Cognitive Reflection Test (CRT) \cite{Frederick} is a widely used measure for the propensity to override System 1 by System 2. It is composed of three-item set of questions designed to initially draw the problem solver into activating System 1. Considering the bat-and-ball problem in the CRT, for instance, using System 1, the tendency is to answer 10 cents. A quick reflection however, indicates that this cannot be correct; for then, the total cost would be $\$0.10 + \$1.10 = \$1.20$! The correct answer using short algebraic manipulation or trial-and-error is 5 cents.

Scores in the CRT are found to have moderate correlation with performance on Wonderlic Personnel Test \cite{Frederick}, heuristics-and-biases tasks \cite{Toplak}, incidences of conjunction fallacy \cite{Oechssler,Liberali}, and susceptibility to some behavioral biases \cite{Hoppe}, amongst others. From a simplified perspective, these studies suggest that people who tend to think deeply, as may be measured by CRT, are able to make better decisions and solutions to problems. In the field of Physics Education, to which this paper belongs, such a perspective although simplified, offers an attractive avenue to look into the possible correlation between CRT and student performance on some standardized tests. The idea is that DPT, as it relates to CRT, may be able to explain the way students approach and solve problems in Physics. This in turn, may give us an insight about the Physics learning process and as educators, the ways by which we may be able to improve it. 

Following similar line of thinking, Wood, Galloway, and Hardy \cite{Wood} (see also Refs. \cite{Kryjevskaia,Gette}) investigated the question ``Can dual processing theory explain physics students' performance on the Force Concept Inventory? [FCI]'' They examined the relationship between student performance on CRT and FCI \cite{FCI,Savi} and found a moderately positive linear correlation between the two for both pretest and post test. The ``findings indicate that students who are more likely to override the system 1 intuitive response and to engage in the more demanding cognitive reflection needed to answer the CRT question correctly are also more likely to score highly on the FCI, implying that similar cognitive processes account for at least some of the cognitive abilities needed for each test.'' \cite{Wood} 

\begin{figure}[htb!]
    \centering
    \includegraphics[scale=1.0]{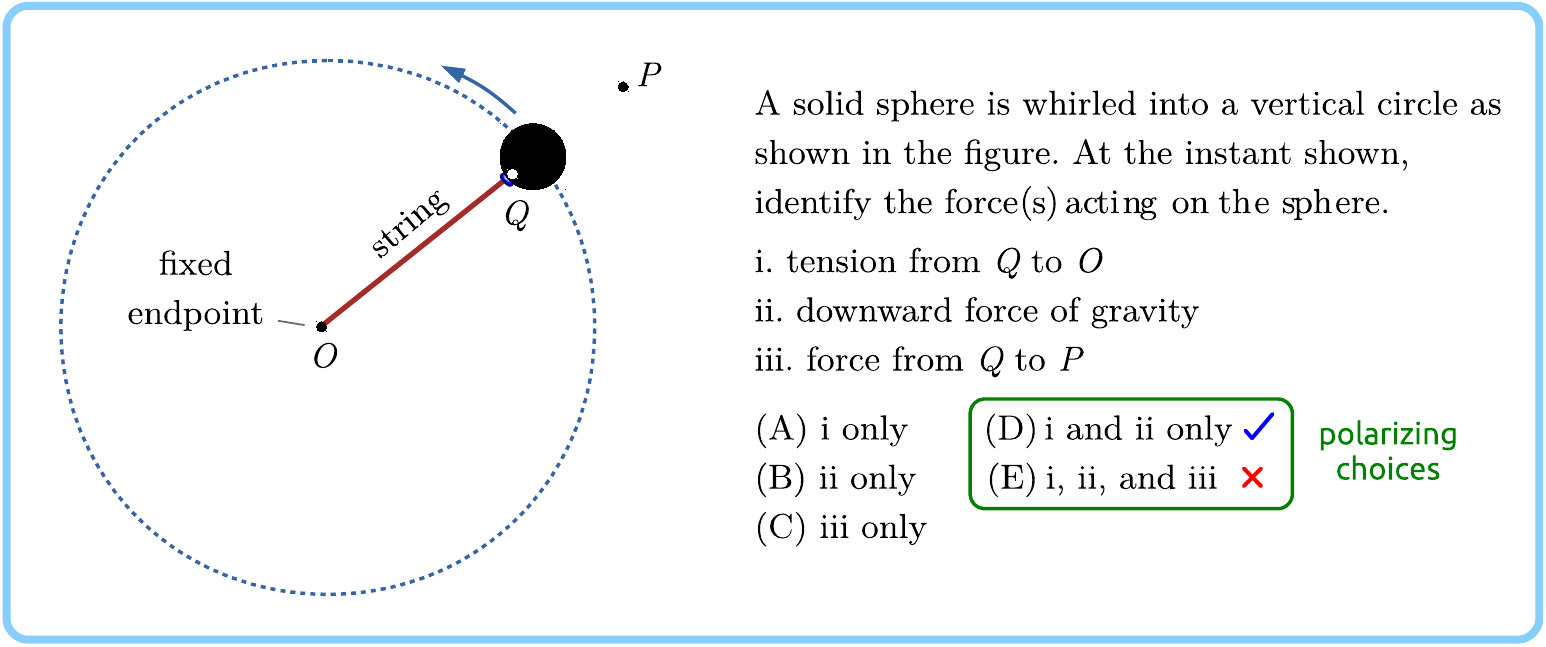}
    \caption{\it In this sample question, one is asked to identify the force(s) acting on the sphere. Whereas most students may be able to identify forces i and ii correctly, the addition of erroneous force iii may confused them, causing ``polarization'' of student answers between choices (D) and (E).}
    \label{fig_1}
\end{figure}

This study intends to further probe the possible relationship between CRT, DPT, and the FCI. In particular, while not necessarily neglecting FCI as a whole, we wish to investigate student performance in the set of six Polarizing Force-Identification (PFI) questions (see Refs. \cite{Alinea, Alinea2}) in the FCI, namely, questions  5, 11, 13, 18, 29 and 30, with their scores in the CRT. This subset of the FCI effectively asks only one basic question: identification of the force(s) acting on a given body. Surprisingly, out of the five choices for each question, the majority of the students tend to only select two choices (polarizing choices). One of these choices is the correct answer containing the right set of forces acting on a given body while the other one contains the right set of forces plus at least one erroneous force; see \textbf{Fig.} \ref{fig_1} for an illustration\footnote{This question is not part of the actual FCI used in this study. It is presented here to illustrate the underlying idea (within the context of this work) about PFI questions. If it may be of help to limit the proliferation of FCI questions for easy student access, we do not quote here actual FCI questions.}. The inclusion of this erroneous force confuses students causing the ``polarization''\footnote{The word polarization is inspired by charge polarization in Physics. A charged body brought near a conducting rod will make one end of this rod positively charged while the other end negatively charged. In a sense, the set of PFI questions ``induces'' this polarization in student response: positive for right answer and negative for wrong answer in the pair of ``polarizing'' choices.} of student response.
\end{multicols}

\begin{multicols}{2}
Dual process theory is especially appealing because of the wide gamut of its applicability from decision-making process in dealing with usual day-to-day encounters to the way we address challenging problems of academic nature. To the extent that CRT could be a link between dual process theory as a possible good measure of the inclination to shift from System 1 to System 2, and the process of learning Physics remains to be further investigated. The pioneering study for the investigation of this link for DPT, CRT, and FCI in Ref. \cite{Wood} is just a beginning. While this work is not a replica of the mentioned study, as part of this paper, we wish to further examine the interconnections among DPT, CRT, and FCI as a whole. In addition to this, we want to probe the set of PFI questions in the FCI, first identified in Ref. \cite{Alinea}, as it relates to CRT and DPT.

This paper is organized as follows. In the following section (Sec. \ref{secmethod}), we summarize the methodology on data gathering and simple statistical analysis. Following this is the results and discussion in Sec. \ref{secresults}. This section is divided into two subsections for separate discussions of the results for the FCI as a whole and CRT, and then for the PFI questions and CRT. In the last section, we summarize our findings and suggest future prospects in line with this study.

\section{Methodology}
\label{secmethod}
The subjects of our study were 163 students under engineering, chemistry, and physics degree programs offered by the University of the Philippines Los Ba\~nos. All of these students have already passed the prescribed calculus-based mechanics course and were taking other fundamental physics course under the author of this work,  upon sitting for the FCI and CRT.\footnote{Because our subjects have already passed a mechanics course, we only needed to administer one FCI. This effectively allowed us to avoid some challenges in explaining the possible shift of correlation between scores in CRT and FCI from pretest to post test, and in analyzing the correlation between normalized gain \cite{Hake} and CRT scores. While these concerns are interesting in their own right, we leave it for further studies.} The two tests were administered as pen-and-paper test and students were given incentives to take the tests seriously. Data gathering took place at the University of the Philippines Los Ba\~nos from Academic Years 2017-18 to 2018-2019.

To minimize possible bias with the CRT scores, we opted to remove results for students who did not take the test for the first time.\footnote{There was no deduction in the incentives provided for students with discarded results.} The remaining 163 test results\footnote{Interested readers may see Ref. \cite{Bialek} for a study on the robustness of the correlation of CRT scores with different variables with multiple exposure to the test and Ref. \cite{Garza} for the effect of \textit{visibility} on the performance in CRT} for the FCI and CRT were subjected to item analysis with the help of ZipGrade \cite{zipgrade} and a spreadsheet application. Common statistical parameters such as mean, standard error, and Pearson coefficients were calculated to find possible meaningful relationships between scores in FCI as a whole, set of PFI questions, and CRT.

\section{Results and Discussion}
\label{secresults}

\subsection{FCI and CRT}
\label{subsecfcicrt}
Table \ref{table_crt} shows the distribution of students with respect to CRT scores ranging from zero out of three (all wrong answers) to three out of three (perfect score). About half (47\%) of the students got a perfect score in the test while only one tenth (10\%) got the lowest score. With 43\% scoring one or two, the skewed frequency distribution drives the average to 2.1/3 (closed to that of Ref. \cite{Wood}, 2.3, with similar cohorts). This is about one and a half times higher than that in Ref. \cite{Garza} covering more than 40,000 students. The difference may be due to stronger inclination of our subjects to mathematics as suggested by their chosen degree programs in engineering, chemistry, and physics, compared to the population with more varied interests investigated in Ref. \cite{Garza}. In Ref. \cite{Sinayev}, the authors found a moderately positive correlation between cognitive reflection and numeric ability.

Majority of the students who got the correct answers perform some sort of short  mathematical calculations on paper. This constitute the majority of our test subjects based on Table \ref{table_crt2}. For the bat-and-ball problem, for instance, majority of the nearly 80\% who got the correct answer, approached the problem by solving some form of algebraic equation(s); some did try trial-and-error to fit the conditions of the problem. For the lilypad problem, geometric sequence either from day 1 to day 48 or backwards from day 48, can be found in the student solutions with correct answers. For the machine problem, short solutions involving the use of ratio and proportion can be found on student papers. Admittedly, few students who got the wrong answers made some marks on paper in addition to the final answer. However, the ``seriousness'' of these scribbles are not at par with those of students who got the correct answer(s). In the lilypad problem for instance, one can find `48/2' yielding 24 days---the intuitive but wrong answer.
\end{multicols}

\begin{table}[t]
\centering
\begin{tabular}{lccccl}
\hline\hline
\small CRT Score                          & \small 0/3         & \small 1/3         & \small 2/3         & \small 3/3         & \small Mean              
\\
\hline
\small Percent Students (No. of Students) & \small 10\% (16) & \small 22\% (36) & \small 21\% (34) & \small 47\% (77) & \multicolumn{1}{c}{\small 2.1} \\ 
\hline\hline
\end{tabular}
\caption{Percent (and number) of students who got CRT scores from 0/3 (all wrong) to 3/3 (perfect score).}
\label{table_crt}
\end{table}

\bigskip

\begin{table}[t]
\begin{center}
\begin{tabular}{lccccl}
\hline\hline
\small CRT Problem                          & \small bat and ball         & \small machine         & \small lilypad                
\\
\hline
\small Percent Students (No. of Students) & \small 79\% (128) & \small 59\% (96) & \small 68\% (111) \\ 
\hline\hline
\end{tabular}
\end{center}
\caption{Percent (and number) of students who answered each CRT \cite{Frederick} question correctly. Problems 1, 2, and 3 are labeled \textit{bat and ball}, \textit{machine}, and \textit{lilypad}, respectively.}
\label{table_crt2}
\end{table}

\begin{multicols}{2}
It is worth noting that students were not required to present a mathematical solution nor any other form of solution to the CRT; only the final answers were required in the closely supervised test. This has been made clear in the instructions explained in class before students could start the tests. Any mathematical solution (be it serious or simply light scribbles) in the form of algebraic manipulation, ratio and proportion, or geometric sequence, is out of their own volition. We find that the majority of students who got the wrong answer in any of the CRT problems simply gave the wrong intuitive answers (i.e., 10 cents, 100 minutes, and 24 days for the bat-and-ball, machine, and lilypad problems respectively) without any supplemental solution at all, or tried to perform some light mathematical scribbles on paper only to end up with or confirm the same wrong intuitive answers; e.g., writing 1:1:1 for the machine problem then yielding the intuitive but wrong answer of 100 minutes. The intuitive answers account for 86\%, 69\%, and 54\% of all the wrong answers in the bat-and-ball, machine, and lilypad problems respectively; this is the majority of wrong answers as in Ref. \cite{Wood}. On the other hand, among the correct student responses are clear cases with erasure marks covering the wrong intuitive answers right beside the correct one. This indicates the transition from System 1 to System 2, consistent with the aim of CRT.

Having said this, there is a possibility that one could have actually activated System 2, as may be evident in their scratch works, for a given CRT question, without finding the correct answer nor the intuitive but wrong answer.\setcounter{footnote}{0}\footnote{This could be avoided by using an online or computer-based administration of CRT. However, some students who are accustomed to doing some scratch works on paper may find some level of discomfort in getting the correct answers even on activation of System 2.} This may then weigh against the accuracy of CRT score as a measure of the tendency to activate System 2. Assuming that students who got the correct answers in the CRT activated System 2 and those who answered the intuitive but wrong answers (10 cents, 100 minutes, 24 hours) activated only System 1, this could possibly account for a maximum of 14\%, 31\%, and 46\% of all the wrong answers in the bat-and-ball, machine, and lilypad problems, respectively. However, consistent with the immediately preceding two paragraphs, we find this group of students to be somewhat of a polar opposite to the group of students who got the correct answer(s)---most students who got the wrong but non-intuitive answers did not write any form of solution at all. Some scribbled one-liner numeric calculation (e.g., `100/5' for the machine problem and `48/4' for the lilypad problem) which may hardly be considered as good evidence of activating System 2. For the three CRT problems we find only three cases (two for the bat-and-ball problem, one for the machine problem, and zero for the lilypad problem) with some tractable short algebraic calculation corresponding to wrong and non-intuitive answers. This gives us confidence to hold on to the raw CRT score as our measure of the level of cognitive reflection at least as far the scope of this study is concerned. 

\begin{figure}[htb!]
    \centering
    \includegraphics[scale=1.0]{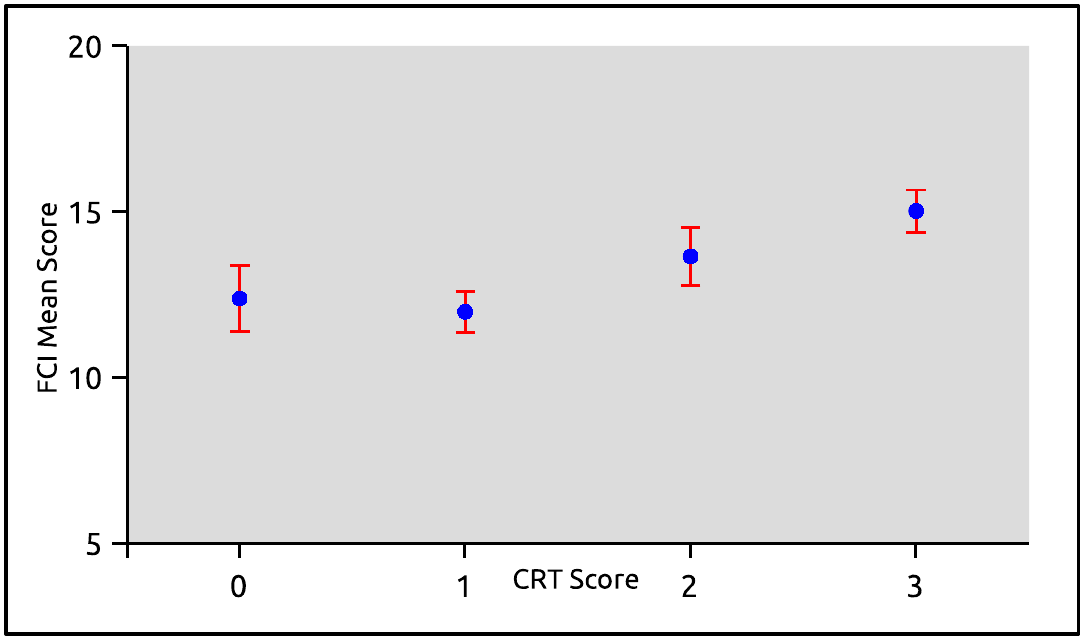}
    \caption{\it Distribution of FCI scores with respect to CRT scores. The error bars are set one standard error of the mean above and below the mean.}
    \label{fig_2}
\end{figure}

After having discussed the CRT results, let us now look into the student performance in the FCI with respect to CRT. Figure \ref{fig_2} shows the distribution of student scores for the two tests. For CRT scores ranging from 1 to 3, the FCI mean score exhibits an upward trend suggestive of a good linear correlation with the former for the mentioned range. However, the FCI mean score for the CRT score of zero, in addition to the errors associated with the mean, somewhat smears this prospect for a possible good positive linear correlation. The Pearson coefficient for the CRT-FCI scores turns out to be 0.24 in the approximate range [0.09, 0.38]\footnote{It is desirable to have this range of \textit{r} as narrow as possible. We approximated that given the same base value of the correlation coefficient, we may need a sample size of $\gtrsim 5000$.} (computed through the Fisher transformation) at 95\% confidence interval. Assuming the same normality condition, this is consistent with the results in Ref. \cite{Wood} for the pretest, $r = 0.38$ in the approximate range [0.23, 0.51] and post test, $r = 0.32$ in the approximate range [0.17, 0.46]\footnote{In the calculation of the approximate interval for the correlation coefficient in Ref. \cite{Wood}, we made used of the based values of this coefficient and the sample size 148 students provided in the reference.}, at the same confidence interval; see Ref. \cite{pearsoncomp} about comparing correlation coefficients.

We provide a two-fold explanation for this finding. Firstly, if CRT is a good measure of the tendency to shift from System 1 to System 2, within the context of DPT, it means that we cannot rule out the possibility that one of the contributing factors in gaining a high score in the FCI is the high level of cognitive reflection. The questions in the FCI each have four out of five incorrect answers in the choices, many of which invoke student misconception or misleading preconception \cite{halloun,finegold} in fundamental classical mechanics. Many of these misconceptions or preconceptions in turn, are from ``common-sense'' everyday experiences that constitute their intuition. For instance, students may have a predisposition to decide right away from ``common sense'' that heavier objects tend to fall faster than a lighter object of the same size and shape. A low level of cognitive reflection can drive a student to select intuitive answers that are wrong. This may then lead, as one of the contributing factors, to low FCI score. 

Secondly, while a shift from System 1 to System 2 may contribute in the more ``serious'' drive to find the correct answer in one FCI question, this may not be enough. After System 2 is activated, some incorrect choices may be discarded after a short but relatively deeper reflection compared to System 1. However, there could remain further hurdles---the other incorrect answers---before a student could pinpoint the correct answer. This means that even if all students start out with activating System 2 without necessarily passing System 1 (similar to that when encountering a math problem to find $\sqrt {365.25}$), they could still end up with the wrong answer, and the student scores could still significantly vary. With the recognition that cognitive reflection as one of the contributing factors in narrowing down the set of correct answers in the FCI, we believe that other factors (e.g., industry measured by study hours, attention span, natural talent, etc.) linked to sufficiently deep understanding of physics also play a significant role in pinpointing the correct answers in the FCI or any other physics tests for that matter. 
\end{multicols}

\begin{multicols}{2}
\subsection{PFI Questions and CRT}
\label{polcrt}
\begin{figure}[htb!]
    \centering
    \includegraphics[scale=1.0]{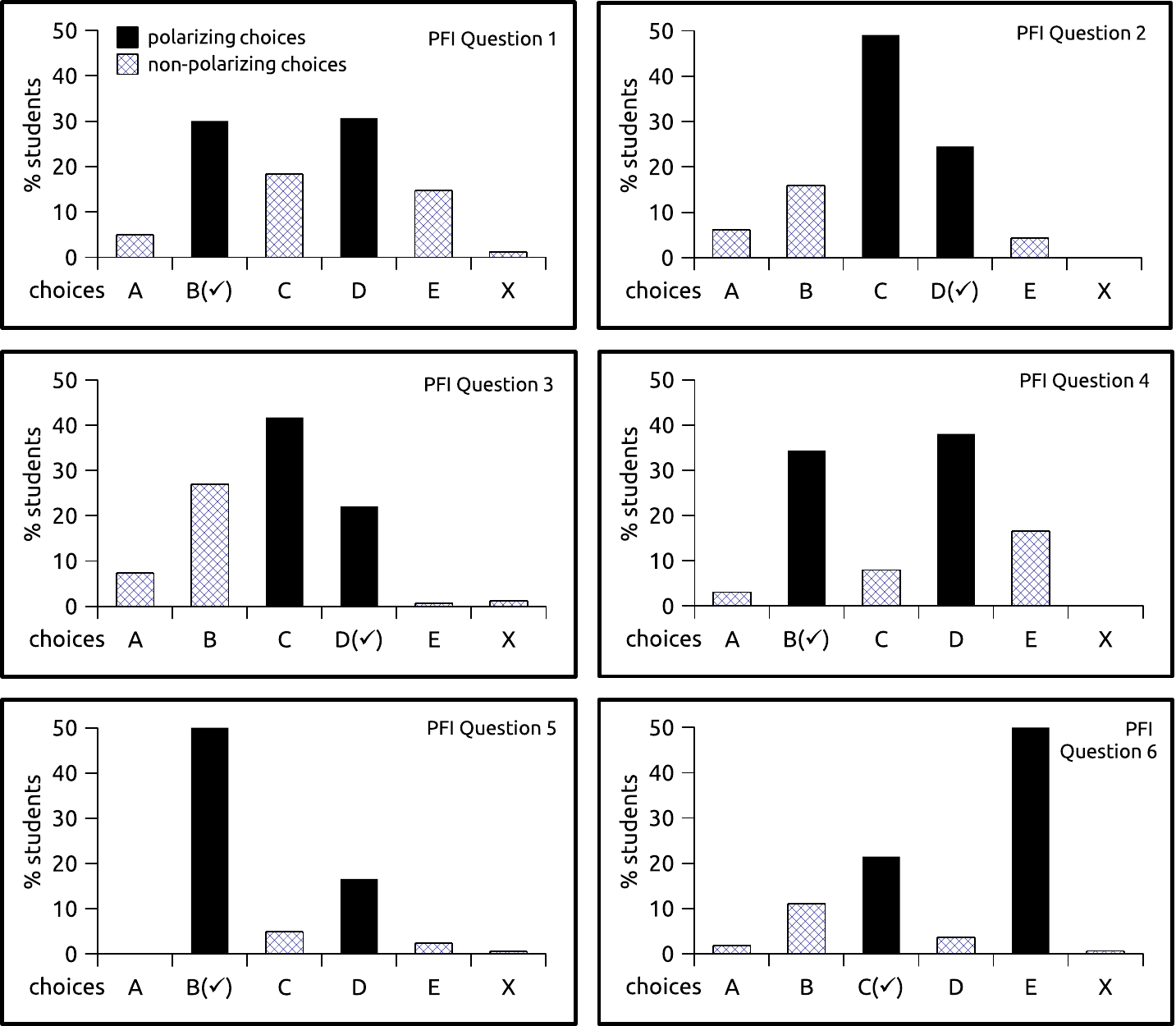}
    \caption{\it Distribution of scores for the six PFI questions (Questions 5, 11, 13, 18, 29 and 30 in the FCI) with respect to CRT scores. Of the five choices, with X meaning no answer, majority of the students chose the two polarizing choices---one that contains the correct answer and the other one that is effectively a superset of the correct answer.}
    \label{fig_3}
\end{figure}

We identified and elaborated the six PFI questions in the FCI in our earlier study presented in Ref. \cite{Alinea}. Back then our sample size was only about 50 (international) students. And although relatively small, the pattern for the PFI questions was sharp enough worthy of publication. Figure \ref{fig_3} confirms the existence of these PFI questions, now with a much larger sample size of 163 students---triple that of the former study. As can be seen in the figure, majority of the students, for the six force identification questions, effectively answer only two (polarizing choices) out of the five choices. Except possibly for the identified PFI question 3, where the number of students who chose letter B is similar\footnote{Further study with larger sample size should resolve the difference between the number of students who elected choices B and D.} to that for letter D (the correct answer), the six bar graphs are consistent with the result in Ref. \cite{Alinea}.
\end{multicols}

\begin{figure}[htb!]
    \centering
    \includegraphics[scale=1.0]{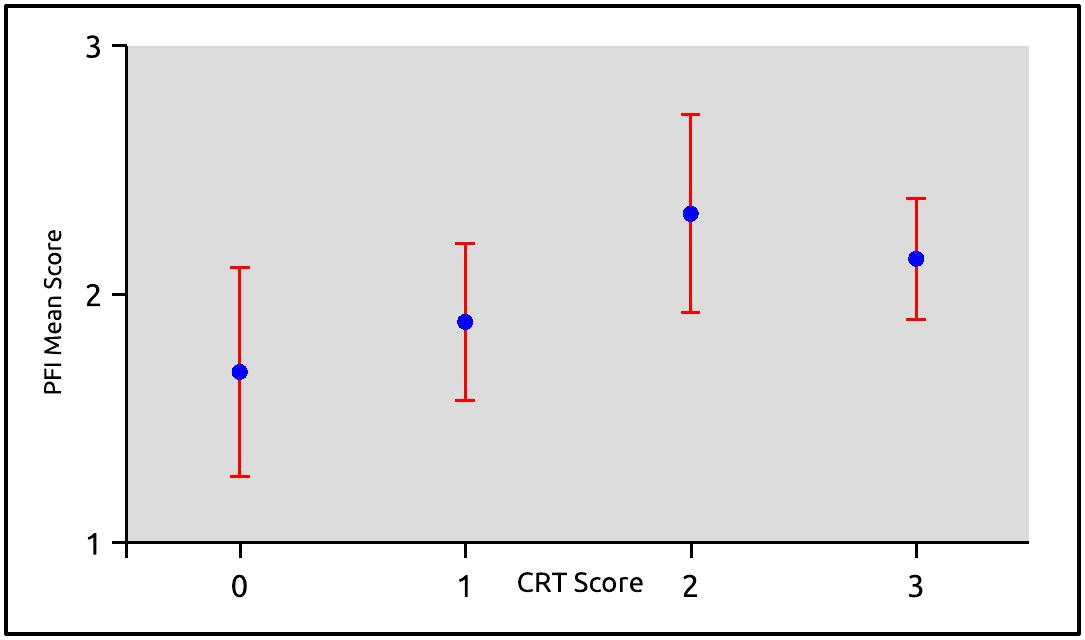}
    \caption{\it Distribution of scores for the set of PFI questions in the FCI with respect to CRT scores. The error bars are set one standard error of the mean above and below the mean.}
    \label{fig_4}
\end{figure}

\begin{multicols}{2}
With the PFI questions at hand, let us look into its possible relationship with the CRT. Figure \ref{fig_4} shows the distribution of student scores for the set of PFI questions with respect to the CRT score. The graph shown is effectively a subset of that shown in Fig. \ref{fig_2} involving all the questions in the FCI. For the PFI questions with respect to CRT, the error bars are wider. There seems to be a rising trend between PFI mean score and CRT score from CRT score of 0 up to 2. However, the downshift at CRT score equal to 3 seems to have spoiled this possible trend. All in all, we get a correlation coefficient of $r = 0.096$ in the approximate range [-0.06, 0.25] at 95\% confidence interval. 

The base value of the Pearson coefficient is small but its range tells us that we cannot simply set aside cognitive reflection in view of student performance in the PFI questions. The identification of forces acting on a given body, as simple as it may sound, is one of the most basic skills necessary in the study of dynamics. Yet even the acquisition of this very basic skill is plagued by misconceptions or misleading preconceptions directly or indirectly from everyday experiences that form our intuition. Instances of these concepts include the requirement for a force to sustain the motion of a body (in a vacuum) and the existence of centrifugal force (in an inertial frame). When confronted with questions asking for a set of forces acting on a given body, low level of cognitive reflection may lead to the inclusion of intuitive but misleading forces. Deep thinkers on the other hand, may rule out these wrong forces, effectively narrowing down the set of prospective correct answers. 
\end{multicols}

\begin{multicols}{2}
Having said this, our range of the Pearson coefficient is evidently on the lower half of its absolute spectrum from 0 to 1. Similar to that for the FCI as a whole, we contend that on top of the ability to shift from System 1 to System 2, is the need for deep understanding of Physics concepts or ideas to pinpoint the right set of forces acting on a given body. In other words, we see that high level of cognitive reflection is some sort of initial ``push'' needed to submerge one into a sea of thought, and complemented by will, natural talent, and/or industry, may lead them to acquire the right understanding or decision in solving physics problems be it as basic as force identification or as complex flying a real rocket.

\begin{figure}[htb!]
    \centering
    \includegraphics[scale=0.9]{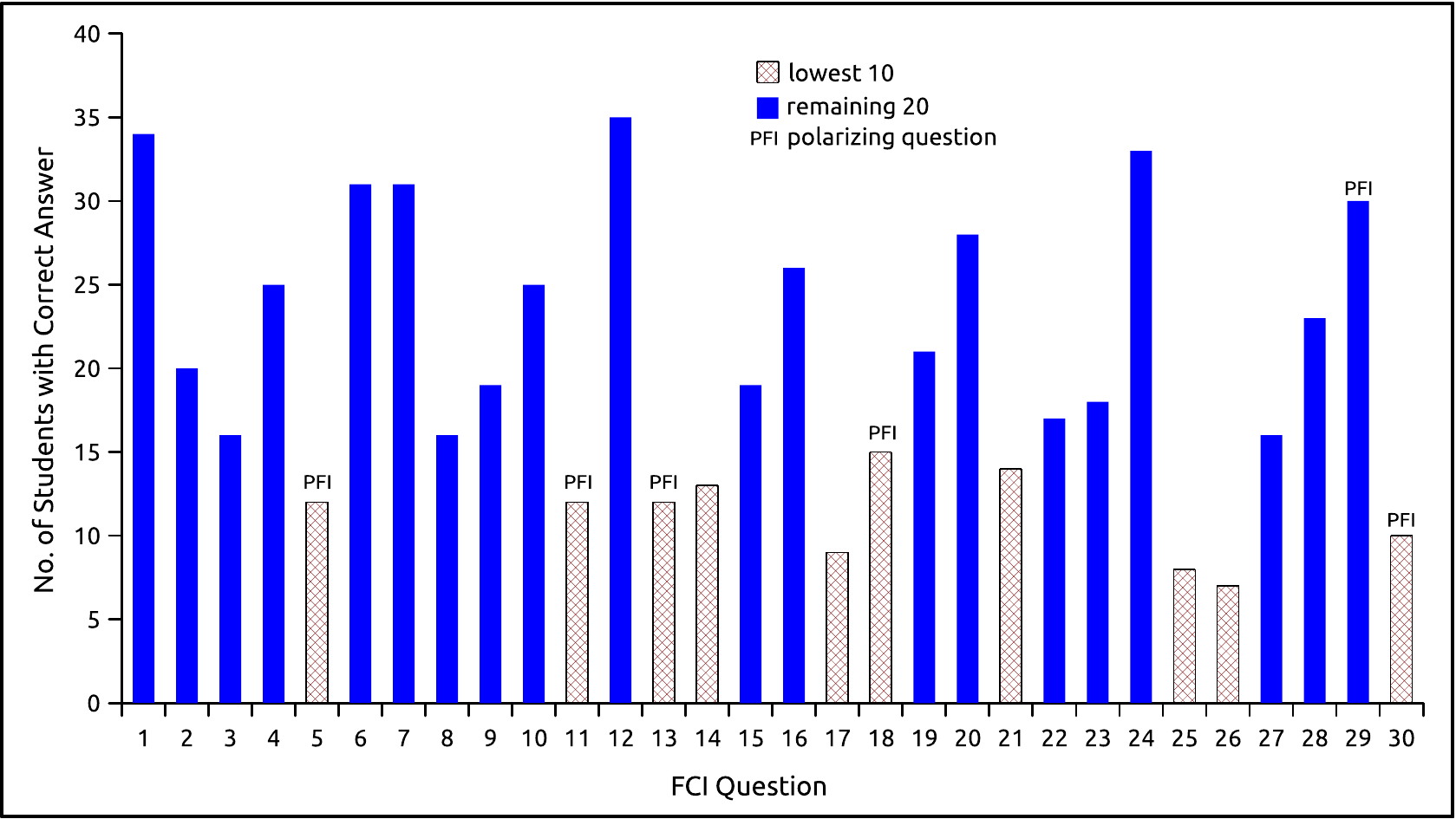}
    \caption{\it Distribution of the number of students who got the correct answer for each question in the FCI. The top 10 (out of 30) most challenging problems based on student performance are identified in graph as the `lowest 10'. Of these 10 questions, five belong to the set of PFI questions.}
    \label{fig_5}
\end{figure}

Before we leave this sub-section, we take cognizance of the downwshifted range for the Pearson coefficient for PFI score vs. CRT score, with respect to FCI score (as a whole) vs. CRT score; that is, from [0.09, 0.38] to [-0.06, 0.25]. Although the intervals are still overlapping, we take our freedom to account for the possible difference (hopefully to be resolved in future studies). Figure \ref{fig_5} shows the distribution of the number of students who got the correct answer for each FCI question. Based on the figure, our students found FCI item numbers 5, 11, 13, 14, 17, 18, 21, 25, 26, and 30, to be the top 10 most challenging FCI questions. Of the six-item set of PFI questions, five belongs to these top 10 questions. We may see then that on average, the subjects of our study found the set of PFI questions to be more challenging compared to the rest of the FCI questions and this possibly caused the downward shift in the correlation coefficient. We are inclined to think that cognitive reflection stands as an important contributing factor in the student performance in the set of PFI questions and in the FCI as a whole. When students activate System 2, this is where the other factors associated with deep understanding of physics come into play. The gravity of these other factors become more evident with the level of difficulty faced by students in answering a given set of physics problems.
\end{multicols}

\begin{multicols}{2}
\section{Concluding Remarks and Future Prospects}
\label{seconclusion}
The operation of the human mind is still one of most complex processes far from our complete comprehension. From the perspective of a Physics Educator, there is a need to understand it so as to optimize student learning experience. But as the goal post of complete understanding is still far beyond the horizon, we are delighted of large strides leading to this end. Dual Process Theory may be seen as one these strides telling us of a significant simplification employing two systems of thought processes: one that is intuitive (System 1) and the other one being analytic (System 2). The Cognitive Reflection Test is a good measuring instrument of cognitive reflection indicating the tendency to shift from System 1 to System 2. The use of CRT offers an attractive avenue to look into the relationship between cognitive reflection and student performance in Physics tests such as the FCI. 

In this study we look into the idea that students who can easily transition from System 1 to System 2 or start right away with System 2 in DPT, may be able to perform better in the set of polarizing force-identification questions in the FCI and in the FCI as a whole. The result of our analysis of tests involving 163 students is suggestive that cognitive reflection is one of the contributing factors involved in student performance in the FCI and the set of PFI questions. Our insight is that high level of cognitive reflection may enable students to cross out seemingly correct choices in the test that are normally part of our intuition from everyday experience or ``common sense'' but otherwise misleading. Complementary to cognitive reflection are other factors associated with deep understanding in physics. These become more evident with the level of difficulty of a given set of physics problems. We find that the range of correlation coefficient between CRT score and PFI score is downshifted with respect to that for CRT score and FCI score. The possible difference may be attributable to the set of PFI questions being more challenging on average compared the rest of FCI questions.

Looking ahead, with all its efforts, this study has only covered a small (but significant) portion of student learning and performance in physics in relation to cognitive reflection and DPT. From here, we foresee future studies involving further tests hopefully with larger number of participants. Other Physics inventories or tests related to scientific reasoning ability (see Refs. \cite{coletta,ates}) may be explored to find correlations with the level of cognitive reflection. Considering CRT, a higher resolution test (see the proposed expanded CRT in Ref. \cite{ToplakME})  in tandem with an expanded PFI questionnaire, may be used to better resolve differences in the level of cognitive reflection. Regarding the administration of the CRT, the use of ``mild subterfuge'' as in Ref. \cite{Wood} may be employed and studied to see its effect on the correlation between cognitive reflection and FCI or any other standardized Physics Test.


\end{multicols}
\end{document}